# Efficient table-top dual-wavelength beamline for ultrafast transient absorption spectroscopy in the soft X-ray region

Lou Barreau[1,2,+], Andrew D. Ross[1,2,+], Samay Garg[1], Peter M. Kraus[1], Daniel M. Neumark[1,2], and Stephen R. Leone[1,2,3,*]

[1]Department of Chemistry, University of California, Berkeley, CA 94720, USA.
[2]Chemical Sciences Division, Lawrence Berkeley National Laboratory, Berkeley, CA 94720, USA.
[3]Department of Physics, University of California, Berkeley, CA 94720, USA.
[*]Corresponding author. Email: srl@berkeley.edu
[+]these authors contributed equally to this work

## ABSTRACT

We present a table-top beamline providing a soft X-ray supercontinuum extending up to 350 eV from high-order harmonic generation with sub-13 fs 1300 nm driving pulses and simultaneous production of sub-5 fs pulses centered at 800 nm. Optimization of the high harmonic generation in a long and dense gas medium yields a photon flux of ~2 x 10$^7$ photons/s/1% bandwidth at 300 eV. The temporal resolution of X-ray transient absorption experiments with this beamline is measured to be 11 fs for 800 nm excitation. This dual-wavelength approach, combined with high flux and high spectral and temporal resolution soft X-ray absorption spectroscopy, is a new route to the study of ultrafast electronic dynamics in carbon-containing molecules and materials at the carbon K-edge.

## Introduction

Transient absorption spectroscopy with high-order harmonic generation (HHG) sources is now established as a powerful technique to probe ultrafast dynamics in atoms[1–3], molecules[4–6] and solids[7–10] (see e.g. ref.[11] for a review). Being an all-optical method, in which the timing is done first and the spectral resolution separately, the time- and energy-resolution are independent and not subject to an uncertainty relationship. It can therefore combine the attosecond temporal capabilities of HHG sources with the atomic-specificity and the chemical environment sensitivity of X-ray spectroscopy. In X-ray absorption spectroscopy, intense absorption features are located around the so-called atomic edges. Therefore, many of the heavier elements show spectroscopic signatures in the extreme ultra-violet region (XUV, 10-100 eV) corresponding to $L_{2,3}$, $M_{4,5}$ or $N_{4,5}$ edges. However, several key elements are not accessible in this spectral domain, including sulfur ($L_{2,3}$ edge at 163 eV), carbon (K edge at 284 eV), nitrogen (K edge at 410 eV) and oxygen (K edge at 540 eV), that are present in the majority of organic and biochemically-relevant molecules and materials. Recent experimental efforts (see e.g. ref.[12] for a review) have thus focused on increasing the photon energy available from table-top X-ray sources, to reach the soft X-ray domain (SXR, 100-1000 eV). This is achieved by taking advantage of the scaling of the cutoff energy in the HHG process with intensity $I$ and wavelength $\lambda$ of the driving laser field as $E_{cutoff} \propto I\lambda^2$ (ref.[13]). While the first SXR photons produced with a HHG source utilized high intensity 800 nm driving wavelength in a strong ionization regime[14,15], the strategy that is now put forward is the use of longer wavelength drivers in the short-wave infrared (SWIR)[12, 16–18] and mid-infrared[19].

In order to provide a spectral supercontinuum instead of discrete harmonic peaks, as well as allow the production of isolated attosecond pulses and therefore considerably improve the temporal resolution, the use of few-cycle SWIR pulses to drive SXR HHG has recently been developed. Indeed, the 9 fs output of an optical parametric chirped pulse amplifier centered at 1600 nm generated a SXR continuum up to 350 eV in neon[20]. By taking advantage of the polarization gating technique with 12 fs, 1800 nm pulses, and the appropriate choice of metallic filters to compensate for the SXR dispersion[21], an isolated pulse of 53 attoseconds extending above 300 eV was measured[22]. Likewise, more conventional approaches relying on the post-compression of 30-50 fs idler pulses from Ti:sapphire-pumped Optical Parametric Amplifiers (OPA) produced SXR radiation supporting isolated attosecond pulses[23, 24]. All of these studies use few-cycle SWIR pulses in the 1600-1850 nm range, which offer the advantages of (i) passive Carrier-Envelope Phase (CEP) stabilization[25] and (ii) the use of transmission through bulk glass for post-compression[26]. However, compared to shorter wavelengths, they suffer unfavourable scaling of the HHG efficiency with the driving wavelength, $\propto \lambda^{-6}$ at the single-atom level[27, 28]. In those conditions, a single absorption spectrum requires accumulation of the signal over a few minutes[22, 29]. Reduced integration times will be decisive for the application of SXR transient absorption spectroscopy for the study of attosecond chemical dynamics at the S, C or N edges[30–32]. Therefore, a trade-off comprising a continuous spectrum extending above the carbon K-edge with high flux and few-femtosecond temporal resolution would be desirable. Moreover, transient absorption in the XUV typically employs a fraction of energy of the few-cycle visible-near IR pulses used for HHG as a pump via strong-field[4, 6] or single-photon excitation[5, 8–10]. Due to the smaller photon energy in the SWIR, using a similar approach for SXR transient absorption considerably reduces the pumping excitation capabilities. Note however that field-driven carrier motion in a semi-metal[33] and tunnel ionization of a low ionization potential molecule[34] with $\lambda > 1600$ nm few-cycle pulses have recently been observed.

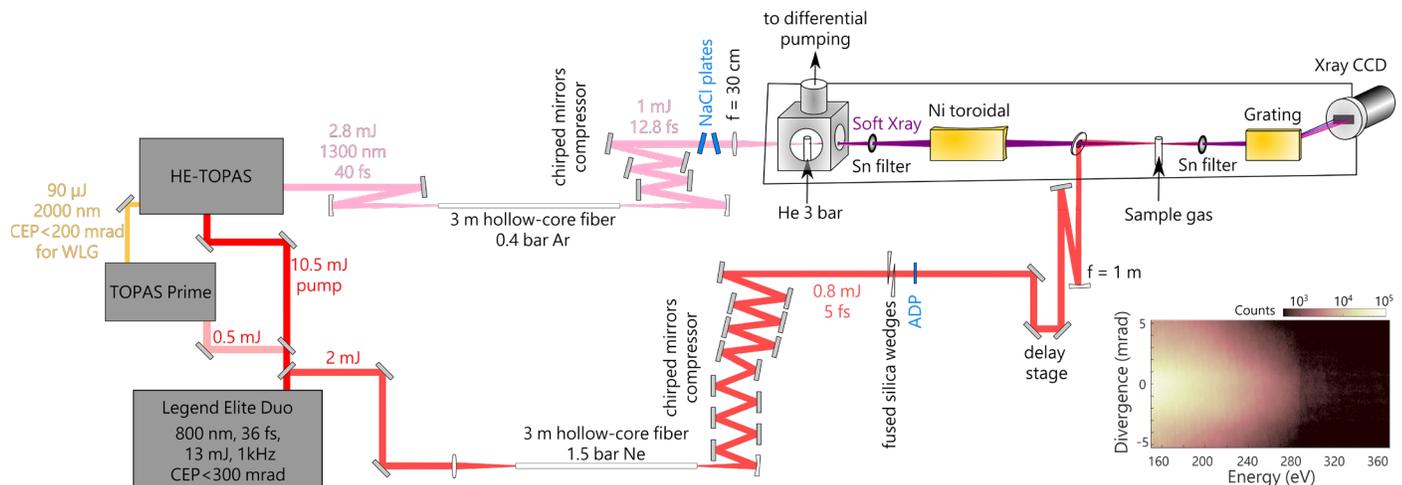

**Figure 1.** Summary of the beamline. Soft X-ray pulses are produced from high-order harmonic generation with few-cycle SWIR pulses in helium. Few-cycle pulses centered at 800 nm are simultaneously produced and are used as a pump in ultrafast transient absorption experiments. The different steps are detailed in the text. A typical X-ray camera image is shown at the bottom right. The drop in signal after ~280 eV is due to carbon contamination of the X-ray optics, causing absorption at the carbon K-edge.

In this work, a table-top beamline simultaneously producing few-cycle SWIR and vis-NIR pulses to be used in ultrafast pump-probe experiments in the SXR is presented. The two-cycle pulses centered at 1300 nm are used to drive HHG in helium, providing a continuous spectrum extending up to 350 eV, with ~2 x $10^7$ photons/s/1% bandwidth at 300 eV. We show that the SXR pulses produced enable the measurement of high resolution absorption spectra in the 170-300 eV range within a few seconds. When used in combination with sub-5 fs vis-NIR pulses, the temporal resolution achieved in a transient absorption experiment at 250 eV is ~11 fs.

## Description of the table-top dual-wavelength beamline

The beamline is summarized in Figure 1. It uses a 13 mJ, 800 nm, 30 fs, 1 kHz Ti:Sapphire laser (Coherent Legend Elite Duo), whose energy is split into 11 mJ +2 mJ to produce the probe and pump pulses, respectively, used in time-resolved X-ray transient absorption experiments. This section details the simultaneous compression of SWIR (centered at 1300 nm or 1400 nm) and visible-near infrared (centered at 800 nm) pulses, and the characteristics of the soft X-ray spectrometer.

### Production of few-cycle short-wave infrared pulses

The few-cycle pulses centered at 1300 nm in the SWIR are produced using an OPA followed by spectral broadening in a hollow-core fiber filled with a rare gas[35] and compression with chirped mirrors[36]. The 11 mJ 800 nm beam is further split into 0.5 mJ and 10.5 mJ for use in a two-stage OPA, which converts the 800 nm to SWIR (1300 nm). The two-stage system is obtained from Light Conversion and is designed to provide CEP stability of the signal pulses. The first stage is a low energy OPA (TOPAS Prime), pumped by 0.5 mJ, which provides 90 $\mu$J of 2 $\mu$m pulses in the idler. Due to the parametric amplification process, the idler pulses are passively CEP stabilized, with a stability of <200 mrad. They are then used as the seed for the white light generation in a second, high energy, OPA (HE-TOPAS) pumped with the remaining 10.5 mJ of 800 nm light. This design should ensure the CEP stability of the signal pulses over the 1200-1600 nm tunable range of the OPA, regardless of the CEP stability of the pump laser. However, the following results are obtained with a CEP-averaged signal pulse. The 2.8 mJ, 1300 nm, 40 fs output is focused with a 2-m focusing mirror to a focal size of 390 $\mu$m (at $1/e^2$) at the entrance of a 3-m-long, 700-$\mu$m-inner diameter stretched hollow-core fiber (HCF) filled with 0.4 bar of argon (Few-Cycle Inc.). The ratio of beam waist to fiber diameter was experimentally adjusted for best transmission through the fiber, and it was found that this slightly smaller value (0.56) compared to the optimal value for coupling to the $LP_{01}$ mode[37] (0.64) yielded higher output power. After broadening through self-phase modulation, the spectrum extends from 1000 nm to 1500 nm (see the dashed line in Figure 2d), supporting 9.3 fs pulses with up to 1.45 mJ per pulse. After recollimation, the pulses are compressed with a set of four ultra-broadband dispersive mirrors (PC1816, Ultrafast Innovations GmbH) that provide an average of -150 $fs^2$ per reflection over the 1000-1700 nm spectral range. For fine-tuning of the group-delay dispersion (GDD), the beam goes through a pair of counter-rotating 2-mm NaCl windows, which have a GDD of +47.8 $fs^2$/mm at 1300 nm. The pulses are characterized using a home-built dispersion-free second harmonic generation Frequency-Resolved Optical Gating (SHG-FROG), whose results are shown in Figure 2a-d. The retrieved pulse has a full width at half maximum (fwhm) in intensity of 12.8 fs, which corresponds to less than 3 optical cycles, with 80% of the energy in the main pulse. Further compression could be achieved with a better control over third-order dispersion.

The ultra-broadband dispersive mirrors are optimized for chirp compensation over a 700-nm bandwidth in the SWIR. In order to test the tunability of the HCF compressor, the central wavelength of the OPA is increased to 1400 nm, the rest of the system remaining identical. The spectrum is broadened in 0.6 bar of argon and extends from 1050 to 1600 nm (Figure 2h), with up to 1.15 mJ per pulse. From the SHG-FROG measurement and retrieval shown in Figure 2e-g, the pulses are compressed down to 9.6 fs fwhm, that is 2 cycles at 1400 nm, with 88% of the energy in the main pulse. Thus, we see that our system allows wavelength tuning in the SWIR over a 100 nm range.

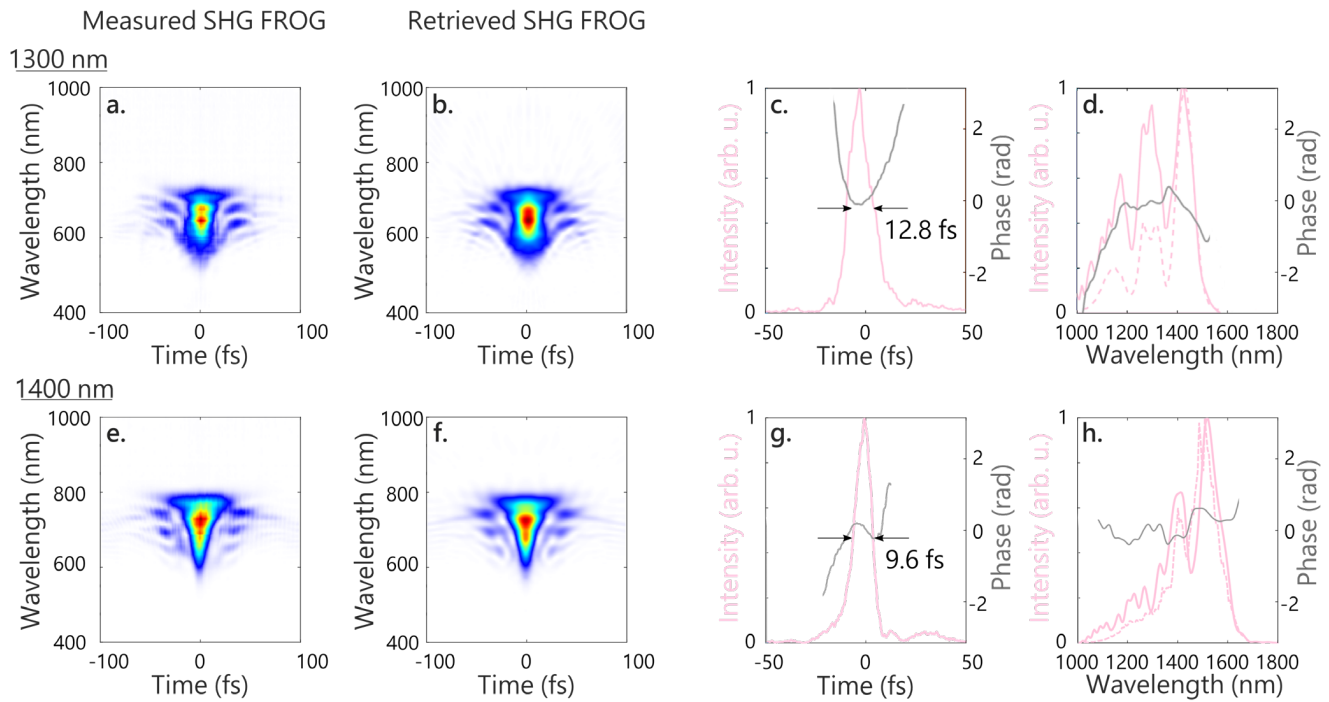

**Figure 2.** Characterization of the compressed SWIR pulses centered at 1300 nm (**a-d**) and 1400 nm (**e-h**) with SHG-FROG. **a,e.** Measured spectrogram. **b,f.** Retrieved spectrogram. **c,g.** Retrieved temporal intensity (pink) and phase (grey). **d,h.** Retrieved spectral intensity (pink) and phase (grey). The dashed line is the spectrum at the output of the fiber as measured with a spectrometer (Ocean Optics FLAME). Phases are shown for intensities >5% of the maximum intensity.

### Simultaneous production of few-cycle visible-near IR pulses

The 2 mJ pulses at 800 nm split before the pumping of the OPA stages (see Figure 1) are compressed to few-cycle duration using the same method[35] in a second, 400-$\mu$m-inner diameter stretched HCF filled with 1.5 bar of neon. After propagation over two meters to compensate for the delay acquired by the SWIR pulse in the OPA stages, the beam is focused with a 2-m focusing lens to a 280 $\mu$m spot size at the entrance to the fiber. At the output, 0.8 mJ per pulse with a spectrum covering the 550-950 nm range (Figure 3c) are obtained. Compression is achieved with reflections on sixteen ultra-broadband dispersive mirrors (PC1332, Ultrafast Innovations GmbH), transmission through fused silica wedges and finally through a 2-mm-thick Ammonium Dihydrogen Phosphate (ADP) crystal, which corrects for third-order dispersion[38]. The visible-near IR (vis-NIR) pulses are characterized with a d-scan module (Sphere Ultrafast Photonics)[39], whose results are shown in Figure 3. The d-scan phase retrieval algorithm indicates a pulse duration of 4.75 fs fwhm, which is 1.1 times the Fourier limit, with 75% of the intensity in the main pulse. The compression of the vis-NIR pulses to 1.8 cycle is realized at the same time as the compression of the SWIR pulses described in the previous section. Therefore, these two pulses can be used together in a pump-probe experiment. We note that shorter pulse durations in the vis-NIR with similar pulse energy can be achieved with this method[38], and could most likely be obtained here if the compression of the Ti:sa laser could be tuned independently for pumping the OPA stages and the vis-NIR HCF compressor.

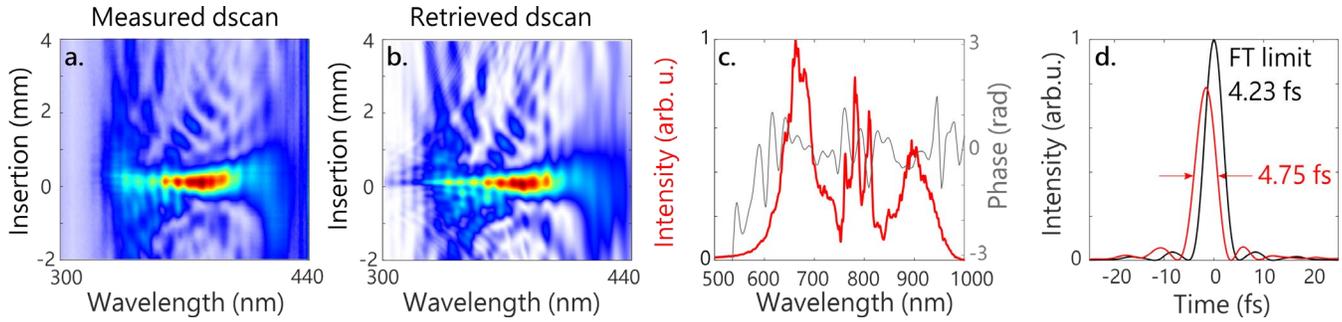

**Figure 3.** Characterization of the few-cycle 800 nm pulses used as a pump. **a.** Measured dscan. **b.** Retrieved dscan. **c.** Measured spectrum (red) and retrieved spectral phase (grey) of the compressed pulses. **d.** Retrieved temporal intensity (red) and Fourier-limited pulse (black).

**Soft X-ray spectrometer**

The soft X-ray (SXR) pulses are produced via high-order harmonic generation by focusing the few-cycle SWIR pulses with a $f$ = 30 cm lens to an intensity of 2 x $10^{15}$ W/cm$^2$ into a 4-mm-long gas cell continuously flowed with helium at a backing pressure of 2.8 bar (see Fig. 1). The use of a lens is unusual with few-cycle driving pulses[5,22–24,40]; however, the GDD near 0 of fused silica for 1300 nm (- 2.4 fs$^2$/mm) provides minimal pulse lengthening from chirp and geometrical sources[41]. The gas cell is surrounded by a 7-cm-edge-length cube for differential pumping, sealed along the propagation axis with cellophane tape through which holes are laser-produced. The cube is evacuated by a roots blower with 900 m$^3$/h pumping speed, such that the pressure inside is calculated to be ~1 Torr and is measured to be ~1x10$^{-3}$ Torr in the surrounding chamber. After HHG, the SWIR light is filtered out with a metallic filter (Zr, Sn, Ag, Ti or Al can be used depending on the spectral region of interest). The SXR is focused by a Ni-coated toroidal mirror at 4° grazing angle into the sample, which can be a gas cell or a solid membrane. Changes between different samples can be easily achieved with motorized stages in the two dimensions perpendicular to the propagation axis. The spectrum is dispersed by an aberration-corrected concave grating and imaged onto an X-ray CCD camera (Greateyes, GE 2048 512 BI UV1). The camera is movable along the grating focal plane to fully capture the SXR spectrum. Depending on the spectral region of interest, two different gratings can be used, optimized for 5-25 nm (50-247 eV, Hitachi 001-0660) or 1-6 nm (207-1239 eV, Hitachi 001-0659), referred to as the low- and high-energy gratings in the following. The grating rotation and camera position can be adjusted to optimize the spectral resolution. The resolution of the spectrometer has been determined using the Ar 2p$^{-1}_{3/2}$ 4s absorption line[42] to be 325 and 190 meV fwhm at 244.39 eV, corresponding to $\Delta E/E$ = 1.3 x $10^{-3}$ and $\Delta E/E$ = 7.8 x $10^{-4}$ for the two gratings, respectively. While the resolution of the high-energy grating is higher, its increased groove density results in a lower efficiency by a factor of ~3. Unless otherwise mentioned, all the spectra shown in the
following are measured using the low-energy grating.

## Optimization of the photon flux at the carbon K-edge

### Driving wavelength

Here we study how the high-harmonic flux in the cutoff region differ for two few-cycle driving pulses of different wavelengths in the SWIR. Indeed, at the single-atom level, the HHG efficiency scales with $\lambda^{-6}$ (refs.[27,28]). However, this goes together with an increase in the cutoff photon energy as $\lambda^2$ (ref.[13]), and calculations have shown that the atomic response can be macroscopically overcome under certain phase-matching conditions[17]. In the previous section, the tunability of the few-cycle SWIR source was demonstrated. We therefore compare the SXR spectra produced when driving the HHG in the 4 mm gas cell with the 1300 nm and 1400 nm pulses characterized in Figure 2. The HHG spectra produced in these conditions are shown in Figure 4a. From the measured spectra, the photon fluxes are calculated at the source using the known quantum efficiency of the CCD camera and tabulated toroidal mirror reflectivity, metallic filter transmission, and grating efficiency. At both wavelengths, the spectra are continuous and extend from below 150 eV (not shown in Figure 4 because of the limited size of the camera chip) to above 300 eV. As expected, the maximum photon energy produced with few-cycle pulses centered at 1400 nm is higher compared to 1300 nm. The photon flux is 2 to 10 times larger at the 1300 nm driving wavelength over the 150-280 eV spectral range. However, in the carbon K-edge region above 280 eV, the two drivers produce similar high harmonic fluxes, around $8 \times 10^5$ photons/s/eV. The measured photon fluxes, in photons/s in 1% and 10% bandwidth at 300 eV, are summarized in Table 1 and compared to literature values for SXR HHG sources driven by few-cycle SWIR pulses. Note that the results from this work, in these generation conditions, are already comparable to measurements with few-cycle drivers of longer central wavelength from HHG in neon[23, 24], being therefore more cost-effective for gases. By changing the HHG gas cell the flux can even be an order of magnitude higher, as will be shown in the following paragraph.

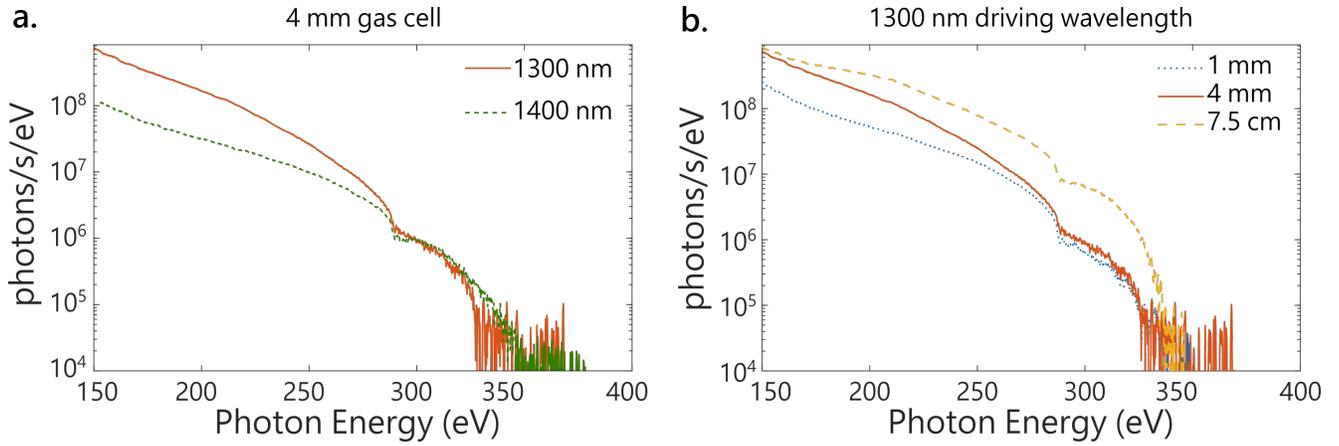

**Figure 4.** Optimization of the flux at 300 eV with the driving wavelength and the gas cell length. **a.** Soft X-ray high-order harmonic spectra generated in helium with few-cycles pulses centered at 1300 nm (red, solid) and 1400 nm (green, dashed) in a 4 mm gas cell. **b.** Soft X-ray high-order harmonic spectra generated in helium with few-cycles pulses centered at 1300 nm in 1 mm (blue, dotted), 4 mm (red, solid) and 7.5 cm (yellow, dashed) length gas cells.

| Condition | Photons/s in 1% BW at 300 eV | Photons/s in 10% BW at 300 eV | Photons/s in 1% BW at 200 eV |
|---|---|---|---|
| 1400 nm, 4 mm cell | $2.8 \times 10^6$ | $2.7 \times 10^7$ | $6.5 \times 10^7$ |
| 1300 nm, 4 mm cell | $2.8 \times 10^6$ | $3.0 \times 10^7$ | $3.5 \times 10^8$ |
| 1300 nm, 1 mm cell | $2.1 \times 10^6$ | $2.2 \times 10^7$ | $1.1 \times 10^8$ |
| 1300 nm, semi-infinite 7.5 cm cell | $2.0 \times 10^7$ | $2.0 \times 10^8$ | $6.7 \times 10^8$ |
| 1850 nm, 12 fs in Ne[23] | | $2.8 \pm 0.1 \times 10^7$ | |
| 1850 nm, 12 fs in He[23] | | $1.8 \pm 0.1 \times 10^6$ | |
| 1800 nm, 12 fs in Ne[24] | $3.0 \pm 0.8 \times 10^6$ | | |
| 1800 nm, 12 fs in He[24] | $1.6 \pm 0.4 \times 10^5$ | | |

**Table 1.** Comparison of photon fluxes in the soft X-ray in existing beamlines using HHG driven by few-cycle SWIR pulses.

**HHG gas cell length**

The question of macroscopic effects in HHG with SWIR pulses has been the subject of recent experimental and theoretical work[16–18, 23]. Here we compare the SXR flux obtained in helium with few-cycle 1300 nm pulses in continuously flowed gas cells of different lengths. The spectrum obtained in a 4 mm gas cell described previously is reproduced in red, solid in Figure 4b. Johnson *et al.* showed that, in the so-called overdriven regime, plasma defocusing effects strongly reshape the HHG driving pulse, already in the tails of the pressure distribution around the gas cell[24]. In order to limit these effects, a shorter gas cell of 1 mm length (1 mm outer diameter and 500 µm inner diameter) is used in place of the 4 mm one described before. The corresponding HHG spectrum, generated in 4 bar of helium, is shown in blue on Figure 4b. No increase of the cutoff is observed, and the flux at 300 eV is 2x less than with a longer interaction medium (Table 1).

Since the absorption length at 300 eV of 2 bar of helium is ~1 cm, a longer HHG medium was also tried[43]. SXR pulses are generated in a 7.5 cm long tube, sealed at the entrance with a 500 µm thick window and with a teflon foil through which a hole is produced with the laser at the end. The focus of the SWIR is towards the exit of the cell, therefore being qualified as *semi-infinite*[44]. No nonlinear effects in the entrance window were observed at the working intensities. A reduced 2 bar helium pressure is required to optimize HHG in the semi-infinite gas cell, so differential pumping is not necessary. Instead, the roots blower pump is used to back the turbomolecular pump attached to the HHG chamber, where the pressure is measured to be $\sim 10^{-3}$ Torr in these conditions. As shown in yellow in Figure 4b, this gas cell produces an order of magnitude more photon flux at 300 eV, with the maximum achievable photon energy remaining identical. Therefore, the semi-infinite gas cell is used in the following experiments. The results call for further investigations of the role of ionization in HHG with few-cycle pulses in a semi-infinite gas cell.

We note that continuous SXR spectra are obtained in all the reported cases, but this observation is not proof of the production of isolated attosecond pulses - these spectra are measured with random CEP and the lack of CEP stability can cause random phase shifts in the HHG spectra that then appear as a continuum[45]. The semi-infinite gas cell yields more photon flux around the carbon K-edge, but is probably less suited for the production of isolated attosecond pulses. However, few-fs SXR pulses at the carbon K-edge would be able to probe ultrafast dynamics on the 5-20 fs timescale, such as passage through conical intersections in organic molecules. Temporal characterization of the SXR pulses[22,46] is needed, in particular in the long gas cell case, but this is out of the scope of the current paper. However, the temporal resolution of the beamline is estimated from a transient absorption experiment in the following section.

## Absorption spectroscopy in the soft X-ray

### Measurement of absorption spectra with the high-harmonic source

In order to demonstrate the suitability of the SXR HHG source, X-ray absorption spectra are typically measured[22, 24, 40, 47–50]. As seen on Figure 1, a gas cell can be inserted in the SXR beam path at the focus of the toroidal mirror. This allows the measurement of spectra such as shown in Figure 5, where the semi-infinite gas cell and 1300 nm pulses are used for HHG. The absorbance, in optical densities (OD), is defined as $A \equiv \log_{10} \frac{I}{I_0}$, where $I$ and $I_0$ are the background-corrected spectra measured with and without the sample, respectively. The high flux of our beamline and the continuous, broad SXR spectrum allow the measurement of absorption spectra in a few seconds only, without scanning the photon energy. The absorption spectrum of $SF_6$ around the sulfur $L_{2,3}$ and $L_1$ edges displayed in Fig. 5a is acquired in 6 seconds (3 seconds with and without the sample). The spin-orbit split doublet corresponding to the excitation of a S $2p$ electron to the unoccupied $a_{1g}$ molecular orbital is well-resolved, as well as a series of excitations to molecular Rydberg states around 180 eV[51]. Above the S $2p^{-1}$ ionization threshold, excitation to quasi-bound states $t_{2g}$ and $e_g$ is visible, as well as the weaker transition of the S $2s$ electron to a $t_{1u}$ molecular orbital around 240 eV. The short acquisition times and the high spectral resolution should allow the observation of ultrafast dynamics in the $SF_6$ molecule[52]. Similarly, absorption spectra of argon around the $L_{2,3}$ edge around 250 eV are obtained in 6 seconds (Figure 5b). At the spectral resolution of 325 meV, three peaks in the series converging to the Ar $2p^{-1}$ threshold can be resolved[29, 42]. Despite the lower flux at higher photon energies (see Figure 4 and Table 1), the absorption spectrum of $CO_2$ around the carbon K-edge can also be measured, this time in 20 seconds (10 seconds with and without the sample). Three main features of the spectrum can be identified, including the main C $1s$ to $\pi^*$ transition[53] (Figure 5c).

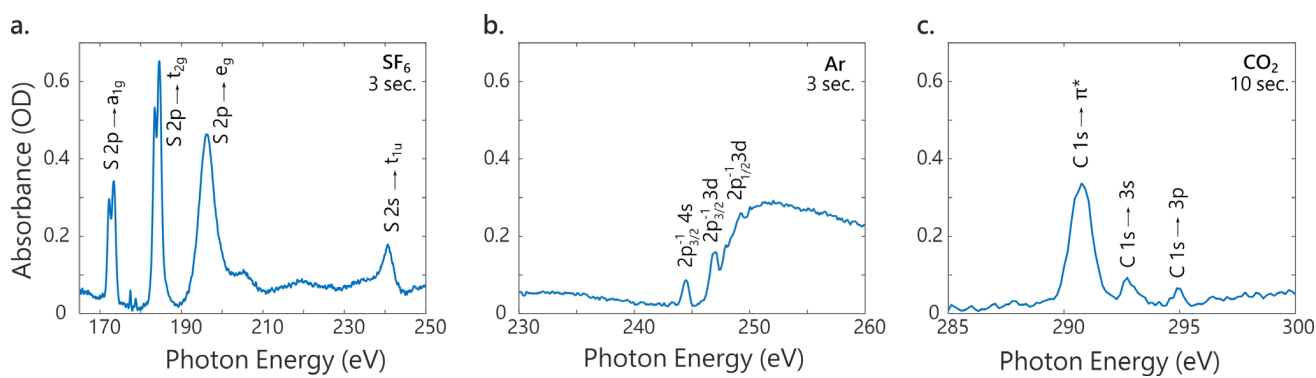

**Figure 5.** X-ray near-edge absorption spectra measured with the SXR HHG source in **a.** $SF_6$, **b.** Ar and **c.** $CO_2$.

### Ultrafast transient absorption in argon near the $L_{2,3}$ edge

In order to realize pump-probe experiments, the sub-5 fs vis-NIR pulse is time-delayed and focused in the sample gas cell with a concave mirror and recombined with the SXR on a hole mirror (see Fig. 1). The optical light is filtered by a metallic foil before the SXR spectrometer. A representative transient absorption experiment is performed in argon, using SXR produced with the 7.5 cm long HHG gas cell filled with 2 bar of helium. The delay-dependent absorption spectra are shown in Figure 6a. At each delay point $\tau$, the absorbance $A$ is calculated as $A(\tau) = -\log_{10} \frac{I(\tau)}{I_0(\tau)}$, where the reference spectrum $I_0$ is reconstructed by using a Fourier low-pass filter directly on the measured signal $I(\tau)$. This procedure allows for removal of the intensity fluctuations of the SXR spectra and is well-suited in the case of narrow absorption lines[3]. In Fig. 6a, $A$ then represents the absorption with respect to the continuum absorption background. Here, the SXR excites 2p core-excited states of argon (Fig. 5b). In the presence of the vis-NIR pulse, with a polarization parallel to the SXR polarization and intensity of 3 x 10$^{13}$ W/cm$^2$, the absorption lines are ac Stark-shifted. This corresponds to a decrease in absorbance at the position of the static lineshape. The spectral resolution (325 meV at this energy, see previous section) prevents the observation of hyperbolic sidebands and the direct measurement of the temporal decay of these states[2, 54]. However, the delay dependence of the pump-induced bleach of the absorption line can give an upper bound of the temporal resolution of our experiment. The changes in absorbance as a function of the SXR-NIR delay for the $2p_{1/2}^{-1}$ 3d, $2p_{3/2}^{-1}$ 3d and $2p_{3/2}^{-1}$ 4s states are shown in Figure 6b-d, respectively. For these three states, the changes can be fit with a Gaussian function of ~11 fs fwhm. We note that in a similar experiment employing SXR and two-cycle 1700 nm pulses, Chew et al. observed sub-cycle oscillations that were attributed to a signature of tunnel ionization[29]. Due to averaging of the CEP, these oscillations are not resolved in our experiments. It would be interesting to study such processes in the future, as the calculated Keldysh parameter in our experiment is also in the tunneling regime but different than at 1700 nm.

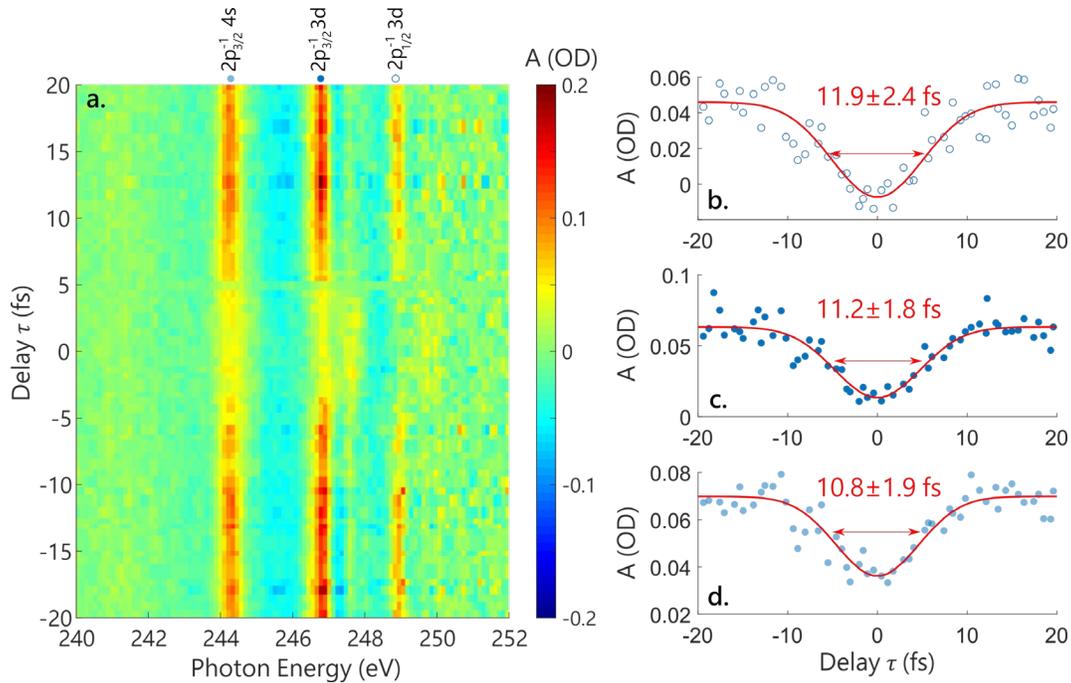

**Figure 6.** Transient absorption spectroscopy in the vincinity of the Ar L$_{2,3}$ edge with sub-5 fs vis-NIR pulses. **a.** Soft X-ray absorption spectra of argon as a function of the delay between the SXR and vis-NIR pulses. Negative delay values correspond to the SXR arriving first. **b-d.** Lineouts of (a) at the $2p_{1/2}^{-1}$ 3d energy (b), $2p_{3/2}^{-1}$ 3d energy (c) and $2p_{3/2}^{-1}$ 4s energy (d). The blue dots are the experimental data and the red line is a Gaussian fit, whose full width at half-maximum duration is indicated.

## Discussion

The simultaneous production of millijoule-level sub-13 fs pulses at 1300 nm and sub-5 fs pulses at 800 nm with two parallel hollow-core fiber compressors has been demonstrated. The few-cycle SWIR pulses can produce a high-flux high-harmonic supercontinuum in the soft X-ray region, with 6.7x10$^8$ photons/s in 1% bandwidth at 200 eV and 2x10$^7$ photons/s in 1% bandwidth at 300 eV. This brightness allows measurement of absorption spectra in the gas phase in a few seconds, an order of magnitude improvement compared to existing SXR HHG sources based on few-cycle driving pulses of longer wavelength[22, 29, 47]. The demonstration of the increased photon flux using a short SWIR wavelength and a long HHG medium represents an important step towards the observation of attosecond dynamics in solution phase with transient absorption in the water window[55]. The dual-wavelength approach furthermore allows the strong-field excitation of molecules[4, 34, 38] or single-photon pumping across a majority of semiconductor bandgaps[8, 9], including 2D semiconductors[56]. The broadband, few-cycle vis-NIR pulse could additionally be upconverted to few-fs tunable UV pulses[57], opening new possibilities for the study of ultrafast excited-state dynamics in molecules[58–60]. A temporal resolution of 11 fs is measured in a pump-probe experiment. This duration is already enough to consider the observation of ultrafast chemical dynamics at the C K-edge[30, 31, 52, 58]. Improved CEP stability and feedback control over the delay will allow attosecond transient absorption experiments in the SXR in the future.

## Acknowledgements

We acknowledge Profs. Roger Falcone, Eric Neuscamann, Feng Wang and Norman Yao for fruitful discussions. We thank Dr. Vincent Gruson and Dr. Stefan Haessler for helpful discussions about the pulse compression, and Dr. Romain Géneaux for comments on the manuscript. This work is supported by the W.M. Keck Foundation Grant No. 042982, the National Science Foundation through NSF MRI 1624322 and matching funds from the Lawrence Berkeley National Laboratory, the College of Chemistry, the Department of Physics, and the Vice Chancellor for Research at UC Berkeley, and the US Department of Energy, Office of Science, Office of Basic Energy Sciences, Chemical Sciences, Geosciences and Biosciences Dividsion under contract DE-AC02-05CH11231, FWP no. CHPHYS01. L.B. acknowledges support from the Miller Institute for Basic Research in Science at UC Berkeley.

## Author contributions statement

L.B. and A.D.R. contributed equally to this work. A.D.R, P.M.K, D.M.N and S.R.L designed the project. L.B., A.D.R and S.G. performed the experiments. D.M.N and S.R.L supervised the project. L.B. and A.D.R. wrote the manuscript, with inputs from all the authors. All authors reviewed the manuscript.

## Additional information

**Competing interests** The authors declare no competing interest.